\documentclass[10pt, conference, compsocconf]{IEEEtran}
\usepackage{cite}
\ifCLASSINFOpdf
   \usepackage[pdftex]{graphicx}
  \graphicspath{{./}}
  \DeclareGraphicsExtensions{.pdf,.jpeg,.png}
\else
   \usepackage[dvips]{graphicx}
   \graphicspath{{./}}
   \DeclareGraphicsExtensions{.eps}
\fi
\usepackage{url}
\begin{document}
%
% paper title
% can use linebreaks \\ within to get better formatting as desired
\title{Enhancing Twitter Data Analysis with Simple Semantic Filtering: \\
Example in Tracking Influenza-Like Illnesses}

% author names and affiliations
% use a multiple column layout for up to two different
% affiliations

\author{\IEEEauthorblockN{Son Doan, Lucila Ohno-Machado}
\IEEEauthorblockA{Division of Biomedical Informatics  \\
University of California, San Diego, USA\\
Email: \{sondoan,lohnomachado\}@ucsd.edu}
\and
\IEEEauthorblockN{Nigel Collier}
\IEEEauthorblockA{National Institute of Informatics\\
Chiyoda-ku, Tokyo, Japan\\
Email: nigel@nii.ac.jp}
}

% conference papers do not typically use \thanks and this command
% is locked out in conference mode. If really needed, such as for
% the acknowledgment of grants, issue a \IEEEoverridecommandlockouts
% after \documentclass

% for over three affiliations, or if they all won't fit within the width
% of the page, use this alternative format:
% 
%\author{\IEEEauthorblockN{Michael Shell\IEEEauthorrefmark{1},
%Homer Simpson\IEEEauthorrefmark{2},
%James Kirk\IEEEauthorrefmark{3}, 
%Montgomery Scott\IEEEauthorrefmark{3} and
%Eldon Tyrell\IEEEauthorrefmark{4}}
%\IEEEauthorblockA{\IEEEauthorrefmark{1}School of Electrical and Computer Engineering\\
%Georgia Institute of Technology,
%Atlanta, Georgia 30332--0250\\ Email: see http://www.michaelshell.org/contact.html}
%\IEEEauthorblockA{\IEEEauthorrefmark{2}Twentieth Century Fox, Springfield, USA\\
%Email: homer@thesimpsons.com}
%\IEEEauthorblockA{\IEEEauthorrefmark{3}Starfleet Academy, San Francisco, California 96678-2391\\
%Telephone: (800) 555--1212, Fax: (888) 555--1212}
%\IEEEauthorblockA{\IEEEauthorrefmark{4}Tyrell Inc., 123 Replicant Street, Los Angeles, California 90210--4321}}

% use for special paper notices
%\IEEEspecialpapernotice{(Invited Paper)}

% make the title area
\maketitle

\begin{abstract}
Systems that exploit publicly available user generated content such as Twitter messages have been successful in tracking seasonal influenza. We developed a novel filtering method for Influenza-Like-Ilnesses (ILI)-related messages using 587 million messages from Twitter micro-blogs. We first filtered messages based on syndrome keywords from the BioCaster Ontology, an extant knowledge model of laymen's terms. We then filtered the messages according to semantic features such as negation, hashtags, emoticons, humor and geography. The data covered 36 weeks for the US 2009 influenza season from 30th August 2009 to 8th May 2010. Results showed that our system achieved the highest Pearson correlation coefficient of 98.46\% (p-value$<$2.2e-16), an improvement of 3.98\% over the previous state-of-the-art method. The results indicate that simple NLP-based enhancements to existing approaches to mine Twitter data can increase the value of this inexpensive resource.

\end{abstract}

\begin{IEEEkeywords}
Twitter; natural language processing; influenza; social media
\end{IEEEkeywords}

% For peer review papers, you can put extra information on the cover
% page as needed:
% \ifCLASSOPTIONpeerreview
% \begin{center} \bfseries EDICS Category: 3-BBND \end{center}
% \fi
%
% For peerreview papers, this IEEEtran command inserts a page break and
% creates the second title. It will be ignored for other modes.
\IEEEpeerreviewmaketitle

\section{Introduction}
\label{s:intro}

The Internet has already proved to be an important resource in surveillance systems for tracking infectious disease outbreaks, showing that there is an opportunity for low cost time-sensitive sources to be exploited to supplement existing traditional surveillance systems. Twitter is an Internet service that offers a social networking and micro-blogging service that allows its users to send and read messages, called tweets. Tweets are text-based posts up to 140 characters. It was shown that by February 2012 Twitter had over 500 active million users, generating over 430 million tweets and handling over 1.6 billion search queries per day \cite{mediabistro-twitter2012,twitter2011}. Earlier work  showed that  self reports in tweets could be used to predict the 2009 A(H1N1) swine flu pandemic \cite{ritterman09}. Recently, Collier \textit{et al.} \cite{collier10-smbm} analyzed self-protective behavior reports in Twitter and showed that there was a moderately strong Spearman's rho correlation between these reports and WHO/NREVSS laboratory data for A(H1N1) in the USA during the later part of the 2009-2010 influenza season. Twitter was used to track the Influenza-Like-Illnesses (ILI ) rate in two recent studies by Lampos and Cristianini \cite{lampos2010} and Culotta \cite{culotta10towards,cullota10detecting}. The key idea in these two studies was to choose keywords to filter and aggregate influenza-related messages. Lampos and Cristianini chose 73 keywords from 1,560 flu-related terms for ILI tracking in the United Kingdom (UK) and compared to Health Protection Agency lab data with a correlation coefficient of 97\% \cite{lampos2010}. Culotta \cite{cullota10detecting} selected only four ILI-related keywords, {i.e., ``flu'', ``cough'', ``headache'',``sore throat''} and reported a high correlation coefficient of 95\%. More recently, Twitter was reported not only as a tool for calculating the ILI rate but for tracking public concerns about health-related events. Signorini \textit{et al.} used Twitter to track levels of disease activity and public concern in the U.S. during the influenza A H1N1 pandemic \cite{signorini2011}. In order to estimate the ILI rate, they first collected Twitter data within the U.S. containing keywords ``swine", ``flu", ``influenza" or ``h1n1" and then built an ILI estimation model using a support vector machine (SVM). The results showed relative high ILI rates with an average error of 0.28\% for national weekly ILI levels and 0.37\% for regional weekly ILI levels. In order to track public concerns, they added more public concern keywords  such as ``travel", ``trip", ``flight" (for  disease transmission) or ``wash", ``hand", ``hygiene" (for disease countermeasures) or ``guillain", ``infection" (for vaccine side effects). Then they calculated the percentage of observed tweets (tweets including public concern keywords) over influenza-related tweets. The results showed that Twitter messages can be used as a measure of public interest or concern about health-related events. Chew and Eysenbach \cite{chew2010} collected Twitter messages containing keywords ``swine flu", ``swineflu" and ``H1N1" and monitored the use of the terms ``H1N1" versus ``swine flu" during the 2009 H1N1 outbreak and validated Twitter as a tracking system for public attention. They showed that several Twitter activity peaks coincided with major news stories and the results correlated well with H1N1 incidence data. For both these studies, keywords play an important role in analyzing the content of tweets.

These related methods combine heuristics and experimentation and raise two important questions: (1) Is there a systematic method to choose high performance keywords for disease tracking? (2) Can richer semantic information contained in Twitter messages such as negation, hashtags or emoticons help improve disease tracking? We investigated a generic approach to filter tweets using two steps. In the first step, we filtered messages using keywords derived from the BioCaster ontology, which is a multilingual public health terminology designed for event surveillance from news media \cite{collier-coling10}. In the second step, we filtered these messages using semantic features. Given the methodology for choosing keywords from the ontology, we called the first step \textit{knowledge-based filtering} and the second \textit{semantic-based filtering}. We chose ILI as a useful example of how to pursue answers to these questions. We systematically investigated the relationship between Twitter messages and the ILI rate. Season influenza (SI) is of particular public health concern since it results in about three to five million cases of severe illness in the worldwide population, causes 250,000-500,000 deaths per year \cite{WHOFluFact} 
and uses substantial hospital resources. At the same time, there is a need to strengthen influenza surveillance in order to help public health professionals and governments quickly identify the signature of novel pandemic strains like 2009 A(H1N1). Among well-known international and national influenza surveillance systems are the WHO Global Influenza Surveillance Network (FluNet) \cite{WHOFluNet}, the US Outpatient Influenza-like Illness Surveillance Network (ILINet) in the US \cite{cdc-flu}, and the European Influenza Surveillance Network (EISN) \cite{ecdc10}. Such systems, which rely heavily on data resources collected from hospitals and laboratories, have high specificity but often lag 1-2 weeks behind because data need to be collected, processed and verified manually by health professionals \cite{cheng09}. Additionally, these systems have high costs for set up and maintenance. There are several studies targeted at using Internet resources for predicting influenza epidemics. Event-based systems for example are now being widely used for detecting and tracking infectious diseases and new type influenza in particular \cite{hartley10}. Additionally, various studies have used pre-diagnostic signals in self-reports or user searches to track the ILI rate. There is currently no standard definition of ILI but it is generally recognized as fever with symptoms of upper or lower respiratory tract infection. Although many reports of ILI will not actually turn out to be influenza, ILI tracking has been shown to correlate well with diagnostic data for both SI and A(H1N1). Chew used user clicks on search keywords such as ``flu'' or ``flu symptoms'' in advertisements and showed a Pearson correlation of 91\% with the ILI rate in Canada \cite{eysenbach06}. Polgreen \textit{et al.} \cite{Polgreen} showed that a set of queries containing terms ``flu'' or ``influenza'' in the Yahoo! search engine correlated closely with virological and mortality surveillance data over multiple years.  Similarly, Ginsberg \textit{et al.} \cite{ginsberg09} developed the widely used system Google Flu Trends (\url{http://www.google.org/flutrends/}) which uses query logs from users in the Google search engine and reported a high Pearson correlation coefficient of 97\%. 

The distinction between this study and previous work rests on two important aspects: (1) we propose here a knowledge-based method to filter tweets based on an extant and publicly available ontology, and (2) we analyzed the role of a wide range of semantic features in relation to the ILI rate using simple natural language processing (NLP) techniques. 

\section{Methods}

\subsection{Data Sets}
\label{s:dataset}
\begin{figure}[t]
\centering
\includegraphics[scale=0.55]{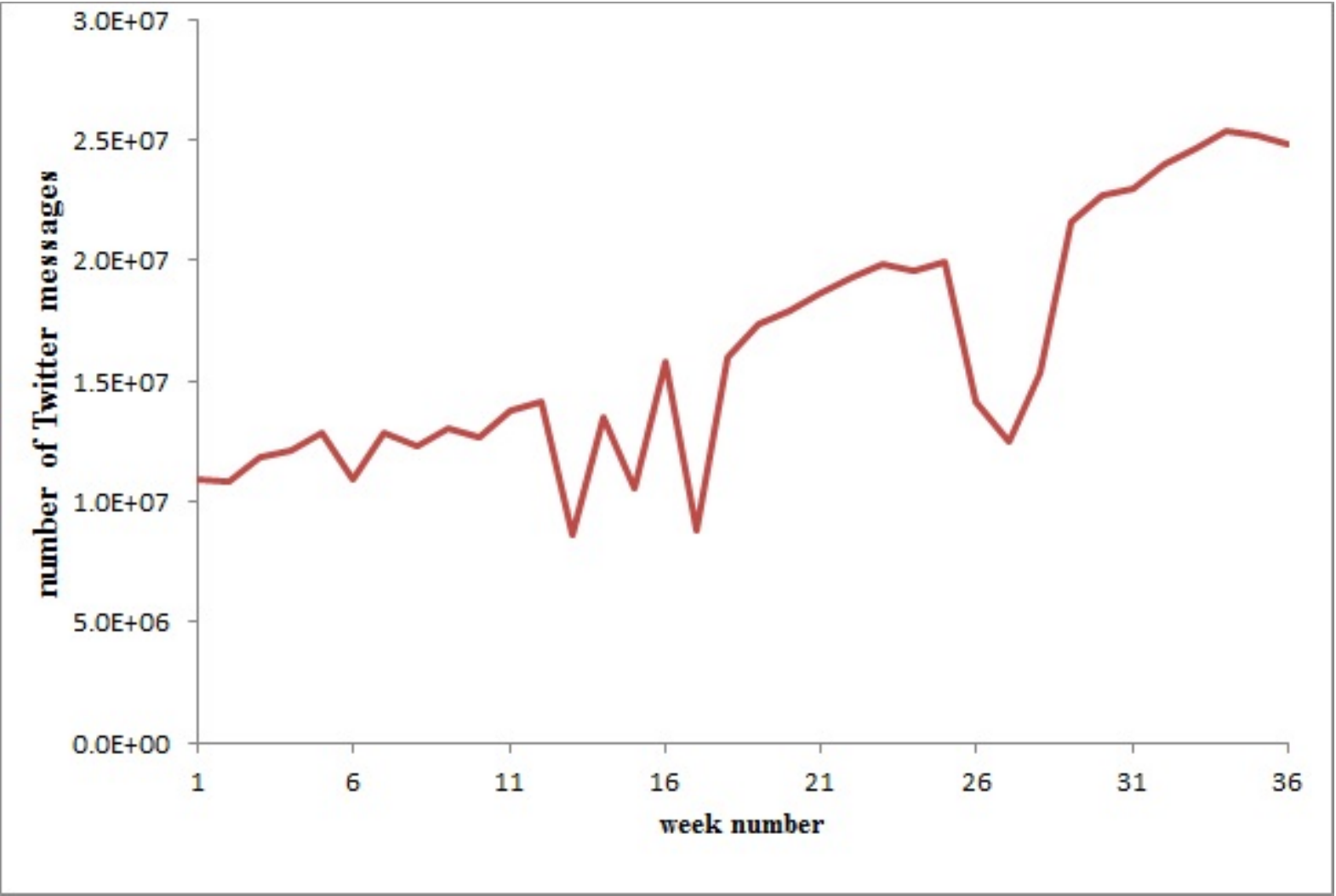}
\caption{Number of Twitter messages per week for 36 weeks. Week 1 ends on September 5th, 2009, week 36 ends on May 8th, 2010.}
\label{fig:tweet-weekly}
\end{figure}

For compatibility and comparison with previous studies, we obtained Culotta's data set from Twitter for 36 weeks, starting from 30th August 2009 to 8th May 2010 \cite{cullota10detecting}. These data were originally collected as part of the work of \cite{brendan10} in which a strong correlation (80\%) was revealed between certain Twitter messages and consumer confidence and political opinion polls in the US using keyword filtering. The obtained data set was selected by ``Gardenhose" real time stream sampling. The total number of tweets over these 36 weeks is over 587 million, containing about 24.5 million unique users. The size of the corpus is about 243 GB in compressed format (.gz files). 
Characteristics of the Twitter corpus are summarized in Table~\ref{tab:1} and the weekly number of tweets is shown in Figure~\ref{fig:tweet-weekly}.

\begin{table}[tb]
\centering
\caption{\textbf{Statistics for the Twitter corpus for 36 weeks. Week 1 ends on September 5th, 2009, week 36 ends on May 8th, 2010, in total of 243GB in compressed file (gz format). URL means a Twitter message containing a URL beginning with the symbol ``http"; Hashtags means the Twitter messages containing tokens starting with the ``\#'' symbol.}}
\begin{tabular}{ccc}
\hline\noalign{\smallskip}
Name     & Unique & Total \\ 
\noalign{\smallskip}\hline\noalign{\smallskip}
Tweets   & -     & 587,290,394 \\ 
Users    & 23,571,765 & - \\ 
URL      & -			 & 136,034,309 \\ 
Hashtags (\#) & 7,963,728  & 96,399,587 \\ 
%Replies (@) & 26,965,397 & 352,185,079 \\
\noalign{\smallskip}\hline
\end{tabular}
\label{tab:1}
\end{table}

Data for the ILI rate from the CDC's U.S. Outpatient Influenza-like Illness Surveillance Network (ILINet) was considered as the gold standard \cite{cdc-flu}. ILINet consists of more than 3,000 healthcare providers in all states of the US, reporting over 25 million patient visits each year. The CDC publishes national and regional ILI rates based on weekly reports from approximately 1,800 outpatient care sites around the US. In this article, the ILI rate for the US was considered as the gold standard so that results could be compared to those in  \cite{cullota10detecting}.

\subsection{Keyword-based Filtering}
\label{s:keywords}
\subsubsection{Empirical approach}
\label{s:empi}

We first used Culotta's method as described in \cite{cullota10detecting}. This method used four keywords ``flu'', ``cough'', ``headache'',``sore throat'' and was reported to have achieved the highest Pearson correlation coefficient of 95\%. We called it \textit{Culotta4} and refer to it as the baseline for comparison. The reason for choosing all four keywords in Culotta's work is that all of them refer to CDC's ILI definition, which is ``fever (temperature of 100 F [37.8C] or greater) and a cough and/or a sore throat in the absence of a known cause other than influenza'' \cite{cdc-flu}. The keyword ``fever" was not chosen since it is highly metaphoric and contained in many irrelevant messages such as ``I've got Bieber fever'' which referred to the pop star Justin Bieber \cite{cullota10detecting}. In addition we used two other filtering methods using keywords: one is described by Signorini \textit{et al.} \cite{signorini2011} which used four keywords ``h1n1", ``swine", ``flu", ``influenza", and the other is described by Chew \cite{eysenbach06} which used three keywords ``h1n1", ``swineflu", ``swine flu". We called the former \textit{Signorini4} and the latter \textit{Chew3}.

While applying keywords filtering is simple, choosing the optimal set of keywords is not trivial. Previous methods were based on a try-and-test strategy, and hence would be difficult to generalize to other diseases. Although our method also relies on expert judgement, it approaches the problem in a systematic manner. In previous work, a set of candidate keywords or queries are selected from a list of pre-defined keywords; second, models are built based on these candidates and compared; finally, the top candidates are chosen based on the highest level of correlation \cite{Polgreen,ginsberg09,lampos2010}. These methods, in our opinion, are complex and require a lot of computation time.  Flu Detector chose 73 keywords from 1,560 flu-related terms on Twitter data \cite{lampos2010}.  Google Flu Trends (GFT) tested each query among over 50 million candidates and finally established the top 45 query terms \cite{ginsberg09}. Recently, Cook \textit{et al.} \cite{cook2011} 
proposed an updated model for GFT and showed that it performed better than GFT during the 2009 H1N1 pandemic. The main difference bewteen these two models is the choice of query terms. The updated model included approximately 160 query terms compared with 45 in the original GFT, with some potential overfitting. The terms in the updated model are more directly related to influenza, especially terms related to influenza symptoms. For example, queries in the categories ``general influenza symptoms" and ``specific influenza symptoms" comprise 69\% of the updated model volume, and 72\% of the updated model queries contain the word ``flu". This result suggested that if we choose more appropriate keywords, we may be able to improve the accuracy of the ILI rate. Moreover, keywords related to influenza symptoms or ``flu" are important to calculate the ILI rate. We propose an alternative approach, not previously tried for this task context, that uses syndrome related terms from an extant ontology. We call this \textit{the knowledge-based approach}.

\subsubsection{Knowledge-based approach}
\label{s:knowledge}

We started from the version 3 of the BioCaster Ontology (BCO)  (\url{http://www.code.google.com/p/biocaster-ontology/}), which is part of the BioCaster project developed by an international team from the National Institute of Informatics, Japan \cite{biocaster}. The BCO contains domain terms in public health such as disease, agents, symptoms, syndromes and species with a focus on laymen's terms that appear in newswire. Syndromes are classified into six categories: dermatological, gastrointestinal, hemorrhagic, musculoskeletal, neurological, and respiratory syndromes. The respiratory syndrome contains 37 symptom keywords as shown in Table~\ref{tab:2}. The simplest method of employing this knowledge is to use all these syndrome keyword for filtering, i.e., we keep all tweets which contain at least one keyword in Table~\ref{tab:2}. This method is called \textit{Syndrome}. We noticed that the respiratory syndrome did not include the keyword ``flu''  which was shown to be an important indicator in the empirical approach since it directly grounds message content to ILI \cite{cullota10detecting,lampos2010} . Thus, we added the keyword ``flu'' into the Syndrome method and called it \textit{Syndrome+``flu''}. 

In addition to technical terms in Table~\ref{tab:2}, we investigated and listed extra terms in Table~\ref{tab:extra}. Basically they are extended from Table~\ref{tab:2} and often used daily in the sublanguage of Twitter messages. These informal terms could be mapped to some ontology concepts. The method use all keywords in both Table~\ref{tab:2} and Table~\ref{tab:extra} for filtering is called  \textit{Syndrome + Extra terms}. The purpose of adding more keywords is that we intended to closely model the types of informal language used in tweets to talk about influenza.

\begin{table*}[tb]
\centering
\caption{\textbf{List of keywords used for respiratory syndrome from the BioCaster Ontology \cite{collier-coling10}.}}
\begin{tabular}{ccc}
\hline\noalign{\smallskip}
achy chest & cold symptom & respiratory failure \\ 
apnea & cough & rhonchi \\
asthma &dyspnea &runny nose\\
asthmatic & dyspnoea  &short of breath \\
blocked nose & gasping for air &shortness of breath \\
breathing difficulties & lung sounds &sinusitis \\
breathing trouble & pneumonia &sore throat \\
bronchitis & pain in the chest & stop breathing \\
chest ache & rales  & stuffy nose \\
chest pain & rales on auscultation &thoracic ache \\
COPD & respiratory arrest & thoracic discomfort \\
chronic obstructive pulmonary disease & respiratory distress & thoracic pain \\
& & tonsillitis \\
\hline\noalign{\smallskip}
\end{tabular}
\label{tab:2}
\end{table*}

\begin{table}[tb]
\centering
\caption{\textbf{List of extra keywords added for respiratory syndrome.}}
{\scriptsize
\begin{tabular}{ccc}
\hline\noalign{\smallskip}
aching chest & flu meds & respiratory distress \\ 
antibiotics & flu paracetomol  & respiratory infection \\
asprin & flu pills  & respiratory problem\\
breath & flu remedy  & respiratory tract infection \\
breathing difficulty  & flu tablets  &sinus  \\
breathing  & flu &sinus asprin \\
chest infection & gasping &sinus ibuprofin\\
chest & gut infection &sinus infection \\
cold asprin & ibuprofen  & sinus paracetomol\\
cold flu & influenza  &sinuses \\
cold ibuprofin & lung & sneezed \\
cold medication  & lung infection & sneezing  \\
cold meds& lymph gland  & sniffled \\
cold paracetomol& nose & sniffling  \\
cold remedy & pain killers & sore thrat \\
cold tablets & painful chest & strep \\
coughed  & paracetomol & swallowing  \\
coughing & pneumonia & throat \\
difficulty breathing & pulmonary & throat infection \\
flu asprin  & respiratory & tonsillitis  \\
flu ibuprofin  & respiratory bacteria infection & tonsils  \\
flu medication & respiratory bacterial infection & whooping cough \\
\noalign{\smallskip}\hline
\end{tabular}}
\label{tab:extra}
\end{table}

Before detailing the semantic features employed in our study, we briefly review related work on categorizing tweets. In an earlier survey, \cite{nardi04} categorized general blogs based on user motivations into five groups: (1) to document user's lives, (2) to provide commentary and opinions, (3) to express deeply felt emotions, (4) to articulate ideas through writing, and (5) to form and maintain community forums. Since Twitter messages are micro-blogs, which are limited to 140 characters, they also have their own unique characteristics.
From the viewpoint of user intensions, \cite{java07} observed that tweets can be classified into four groups: (1) daily chatter, (2) conversations, (3) sharing information or URL, and (4) reporting news. It is easy to see that we can group two first categories in the study of \cite{java07} as \textit{individual reports}, and the remaining as \textit{public reports} or \textit{comments on external events}. We separated them into two such groups because (1) the ILI rate depends on the number of specific ILI cases, i.e., individual reports, and (2) two both categories can be easily recognized using the presence of a URL. 
We assumed that individual reports do not usually contain a URL due to the lack of space, whereas comments on external events often contain URLs within their text. Examples of tweets about external events include: ``7-year-old boy dies of flu, pneumonia $<$URL$>$'' or ``Major swine flu symptoms:Sore throat, prod. cough, runny nose, fever, headache, lethargy - $<$URL$>$ ''. We called the individual reports obtained after  filtering to remove those messages with URLs as \textit{Syndrome+``flu''-URL}.

\subsection{Semantic-based Filtering}
\label{s:semantic}

\begin{figure}[t]
\centering
\includegraphics[scale=0.33]{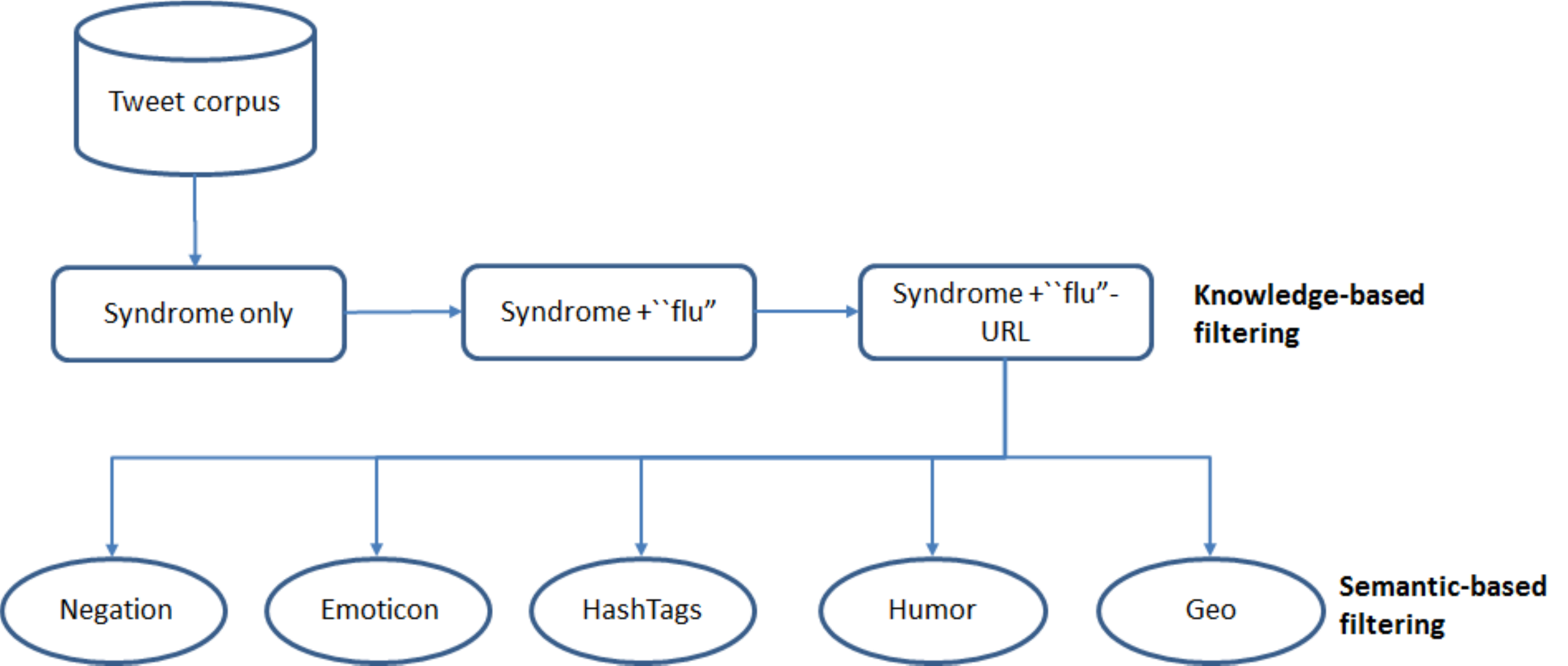}
\caption{A general schema for filtering Twitter data.}
\label{fig:filtering-scheme}
\end{figure}

In order to identify useful features that might help filter Twitter messages, we chose five candidate semantic features: negation, hashtags, emoticon, humor, and geography. Below we explain the reasons for these choices and the way we extracted them from the Twitter messages. Since the ILI rate depends largely on identifying individual cases, we explored those features after filtering for individual reports. A general schema for filtering in this paper is depicted at Figure~\ref{fig:filtering-scheme}.

\begin{figure}[t]
\centering
\includegraphics[scale=0.33]{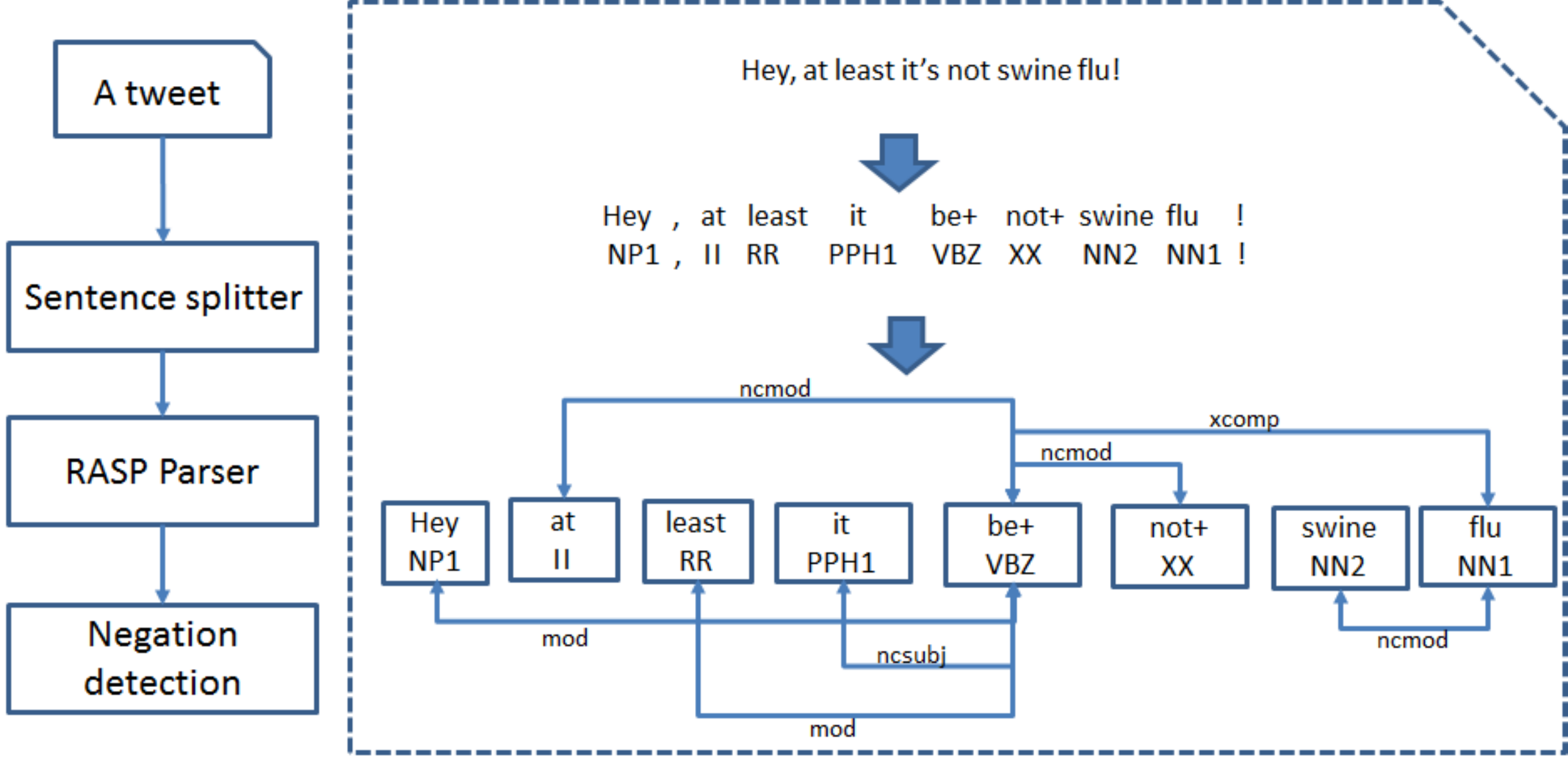}
\caption{The scheme for detecting negation in tweets and an example. In the example, ``NP1", ``NN1", ``NN2", ``VBZ", ``PPH1", ``II", ``RR", ``XX" are tags defined in CLAWS taget \cite{Garside_CLAWS87}, ``mod", ``ncmod", ``ncsubject", ``xcomp" are grammatical relationships defined in the RASP parser \cite{briscoeCW06}. We identify the negation by the following rule: if there is direct or indirect grammatical relationship between the negation tag (``XX") and the term ``flu" , we remove the tweet. In this example, there is an indirect relationship between them through ``be+" (VBZ), therefore, this tweet is removed.}
\label{fig:negation}
\end{figure}

We first discuss the use of negation in tweets, which have a wide range of linguistic functions in English. In general, determining the scope of negation is quite complex and challenging for linguistics in general and sentiment analysis in particular \cite{Wiegand2010}. We were motivated by the simple observation that a negative sentence that talks about users who are not infected by ILI, usually has a negation before the term ``flu''.  For example, ``It's not about having flu'' or ``it's not swine flu'' are negative sentences, but the sentence ``not feeling well, got flu and cough'' is not a negative sentence in this regard. Two processing steps are necessary:  (1) breaking tweets into individual sentences, and (2) parsing each sentence and identifying the negation. To this end, we employed the RASP grammatical parser developed at the Computer Laboratory in Cambridge University \cite{briscoeCW06}. The RASP parser works in a pipeline as follows. First, it takes raw text as the input, runs a sentence boundary detection program to break it into multiple sentences; second, it runs a tokenization program, and each token is then tagged by one of 150 part-of-speech and punctuation labels derived from the CLAWS tagset \cite{Garside_CLAWS87}. Finally, the parser uses a generalized parser to choose the most probable parsing tree based on probabilistic ranking. The parsing tree is then converted into a grammatical relation that determines relationship between subjects and objects in a sentence. 
We identify negation as follows. First we identify whether the tag name for negation (named ``XX" in the CLAWS tagset) is present in a sentence containing the term ``flu". Second, we apply the simple rule: if there is a direct or indirect grammatical relationship between the negation and the term ``flu", then we remove that tweet. The idea behind this rule is if texts such as ``it's not swine flu'' or ``I don't have flu" then they will be removed.
We called this filtering method \textit{Negation}. A scheme and an example for detecting negation are depicted in Figure~\ref{fig:negation}. 

Hashtags and emoticons were investigated previously as part of the filtering problem and showed to be important indicators for improving the accuracy of filtering \cite{reyes10,go09,davidov10}.  A hashtag is a token beginning with the symbol ``\#''. It is considered as a community-driven convention for adding additional context and metadata to tweets. Hashtags were developed as a means to create ``groupings'' on Twitter, helping users to emphasize or group important information or topics, therefore they could be important indicators for the ILI rate. We identified ILI-related hashtags as ones in any sentence containing respiratory syndrome keywords in Table~\ref{tab:2} or the term ``flu''. For example, ``Still coughing smh \#swineflu \#h1n1''. If a tweet contained a hashtag that was not related to ILI, we removed it. We called this filtering based on ILI-related hashtags \textit{HashTags}.

Emoticons are another key feature in tweets that express the mood of users, e.g., anxiety, anger, happiness. We used a list of emoticons from Wikipedia \cite{wiki-emoticon} to identify ``smiley'' emotions of users in sentences. They are emoticons with smiley or laugh, hug, love, heart, or groups of characters such as ``:-)'', ``:)'', ``:D''. For example, ``Glad to hear that you're beating the flu. :-) Hope you don't get the nasty cough that everyone's getting this year''. These smiley icons might not be relevant to ILI. If a tweet contained emoticons with smiley or laugh, hug, love, heart, or shade icons, it was filtered out. We called this method \textit{Emoticon}.

Humor features indicate whether the content of tweets is a joke, irony or humor and could be a strong clue that the tweet is not relevant to ILI. We identified humor features with keywords such as ``haha'', ``hihi'' or ``***cough ... cough***''. We observed lots of tweets containing the phrase ``***cough ... cough***'' are jokes, for example,  ``Hm Im kinda wanting to go to NYC really soon ***cough ... cough*** @Ctmomofsix =)''. We called this filtering method \textit{Humor}.

Geography is an feature associated with Twitter messages that could have particular usefulness in surveillance systems. Such information is not located within the tweet's message line, but is stored optionally in the user's profile data. Since each user is free to enter any information, the geographical information can be a country,  city or town name such as ``NY, USA'', ``LA'', or latitude/longitude from mobile devices such as   ``\textbackslash u00dct: -7.272681,112.755908'', or even nonsense information. Since CDC data cover only the USA, we needed to keep only Twitter messages for this country. To do this, we sent queries about Twitter's users locale to Google Map (\url{http://maps.google.com}) and obtained returned results in JSON format. We wrote a simple parser in Python to parse these returned results to get information about the country. We selected tweets which only came from the USA and designated this filtering method as \textit{Geo}.

\section{Evaluation}
\label{s:eval}

\subsection{Evaluation measures}
\label{s:metric}

We used the Pearson product-moment correlation coefficient to measure the
correlation between fluctuation in the number of tweets according to each of the filtering methods described above and the CDC data on ILI rates. Although not very sophisticated, this measure was used in
previous work to compare standard time series data \cite{eysenbach06,ginsberg09,lampos2010,cullota10detecting,cook2011}, so we adopted for easy comparison of our methods with previously published algorithms. As a reminder to the readers, the Pearson correlation coefficient between two variables X and Y is calculated as follows. 
$$
r=\frac{\sum_{i=1}^{n}(X_i-\bar{X})(Y_i-\bar{Y})}{\sum_{i=1}^{n}{(X_i-\bar{X})^2(Y_i-\bar{Y})^2}}
$$
where $X_i$,$Y_i$ are samples and $\bar{X}$, $\bar{Y}$ are sample means of variables $X$ and $Y$, respectively. The value of r represents the linear relationship between the variables and ranges from -1 to +1. It  is +1 or -1 in the case two variables are perfectly correlated or anti-correlated, respectively. If it is 0, there is no linear correlation between the variables. 

In our experiments, $X$ designates the ILI rate from CDC data, and $Y$ was a normalized value representing the frequency of filtered Twitter messages. Since CDC data are reported weekly (starting on Sunday and ending on Saturday), we bucketed Twitter messages by week. $X_i$ and $Y_i$ were the number for the ILI rate and the normalized number of filtered Twitter messages for the week number i (i = 1, ..., 36), respectively. Statistical analyses were implemented using R packages (\url{http://www.r-project.org/}).

\subsection{Results}
\label{s:result}

The result of different keyword-based filtering methods is shown in Table~\ref{tab:keyword} with all p-value$<$2.2e-16. Among three keyword filtering methods, \textit{Culotta4} had the highest correlation of 94.85\%, which was slightly but not statistically higher than \textit{Signorini4} (94.73\%) and \textit{Chew3} (94.48\%). \textit{Culotta4}'s method kept the highest number of tweets (more than 1.8 million tweets), while \textit{Chew3} just got over 300 thousands tweets. This contrasted with the results from another source, Google Flu Trends, which used search query logs and achieved a correlation coefficient of 99.12\%. 

Table~\ref{tab:keyword} shows that the correlation coefficient of \textit{Culotta4} was 94.48\%, significantly higher than \textit{Syndrome} at 88.60\% (p=0.03). However, when adding the keyword ``flu'' into \textit{Syndrome}, the coefficient increased to 97.13\%, which was significantly higher than \textit{Culotta4} (p=0.0006). This indicated that using keywords derived from the knowledge-based ontology yielded better results when compared to simple syndrome keywords only. The keyword-based methods also showed that removing URLs from tweets helped significantly improve correlations to 97.52\% with \textit{Syndrome + ``flu'' - URL} (p=0.0002). Interestingly, these results supported the results of Cook \textit{et al.} \cite{cook2011} that showed that by considering more ``influenza symptom" keywords to search query logs, the Google Flu Trend models improved. This implies that not only the term ``flu", but symptom keywords are probably important to calculate the ILI rate in both search query logs and Twitter data.

Adding the keyword ``flu" into the filtering method led the correlation to increased significantly (nearly 10\%, p=0.0002), but the number of tweets decreased dramatically. On the other hand, combining extra terms into syndrome terms as in the \textit{Syndromes + Extra terms} method increased the correlation from 88.60\% to 95.78\% (p=0.007), but still kept the number of tweets high more than 880,000 tweets). Compared to \textit{Culotta4}, \textit{Syndromes + Extra terms} had higher correlation but not significant (p=0.29). 

\begin{table*}[tb]
\centering
\caption{\textbf{Results using keyword-based methods in Twitter messages (results for Google Flu Trends (GFT) is shown for reference, since GFT uses search query data, not tweets) (all p-value $<$2.2e-16). Bolded texts show the best correlation coefficient scores.}}
\begin{tabular}{llcc}
\hline\noalign{\smallskip}
& Keyword-based filtering & \ \#tweets &Pearson correlation \\ 
&                         & &coefficient (\%) \\ 
\noalign{\smallskip}\hline\noalign{\smallskip}
Empirical  &\bf{Culotta4 (baseline)} & \bf{1,812,682} & \textbf{94.85} \\ 
approach								&Signorini4 & 1,294,459 & 94.73 \\ 
								&Chew3 & 307,884 & 94.48\\ 
								&Google Flu Trends	& NA & \textit{99.12} \\
\noalign{\smallskip}\hline\noalign{\smallskip}
								&Syndromes only	& 386,199 & 88.60 \\
Knowledge-based  &Syndromes + ``flu'' & 9,034 & 97.13 \\
approach				  &\bf{Syndromes + ``flu'' - URL} 	& \bf{8,485}  & \bf{97.52} \\
											&Syndromes +  Extra terms	& 887,718  & 95.78 \\
											&Syndromes + Extra terms 	- URL & 668,611  & 94.43 \\																			
%											&Syndromes + Extra terms + Geo & 213,989  & 96.03 \\
%											&\bf{Syndromes + Extra terms + Exclude}  & \bf{1,115,885} & \bf{95.70} \\
%											&Syndromes + Extra terms 	+ Exclude - URL & 852,508  & 94.16 \\
\noalign{\smallskip}\hline
\end{tabular}
\label{tab:keyword}
\end{table*}

Results showing the effectiveness of each semantic feature and their combination were shown in Table~\ref{tab:sem} (all p-value $<$2.2e-16). We designated a combination by adding a symbol \textit{``+''} between features. For example, \textit{Negation + Emoticon} means the combination between the filtering methods \textit{Negation} and \textit{Emoticon}. Among the four semantic features, negation proved to be the least effective since it improved  \textit{Syndrome + ``flu''- URL} only from 97.52\% to 97.65\%. The other remaining features showed a slightly positive effect but also not statistically significant improvement. Table~\ref{tab:sem} shows that the effectiveness of semantic features on the ILI rate could be ranked by the following order: Emoticon $<=$ Hashtags $<=$ Humor $<=$ Negation. 

\begin{table*}[tb]
\centering
\caption{\textbf{Results using semantic-based methods with individual reports in Twitter data (all p-value$<$2.2e-16). Bolded texts show the best correlation coefficient score.}}
\begin{tabular}{lcc}
\hline\noalign{\smallskip}
Semantic-based Filtering	& \ \#tweets & Pearson correlation \\ 
                  & & coefficient (\%) \\ 
\noalign{\smallskip}\hline\noalign{\smallskip}
%Baseline (Culotta4) &  1,812,682 & 94.85 \\
%Signorini4 & 1,294,459 & 94.73 \\ 
%Chew3 & 307,884 & 94.48\\ 
%Google Flu Trends	& NA & \textit{99.12} \\ 
%\noalign{\smallskip}\hline\noalign{\smallskip}
Individual reports (Syndromes + ``flu'' - URL) & 8,485 & 97.52 \\
Negation & 7,359 & 97.65 \\
Emoticon & 8,285 & 97.52 \\
HashTags & 8,315 & 97.61 \\
Humor & 8,388 & 97.65 \\
Geo & 2,214 & 98.39 \\
Negation + Emoticon & 7,192 & 97.62 \\
Negation + HashTags & 7,218 & 97.70 \\
Negation + Humor & 7,268 & 97.74 \\
Negation + Emoticon + HashTags + Humor & 6,978 & 97.76 \\
\textbf{Negation + Emoticon + HashTags + Humor + Geo} & \textbf{2,180} & \textbf{98.46} \\
\noalign{\smallskip}\hline
\end{tabular}
\label{tab:sem}
\end{table*}

\begin{table*}[tb]
\centering
\caption{\textbf{Results using \textit{Geo} filtering with \textit{Syndromes + Extra terms} (all p-value$<$2.2e-16).
We only see an improvement by filtering by geography \textit{Geo}, but not for other features (Negation, Emoticon, HashTags, Humor).}}
\begin{tabular}{lcc}
\hline\noalign{\smallskip}
Semantic-based Filtering	& \ \#tweets & Pearson correlation \\ 
                 & & coefficient (\%) \\ 
\noalign{\smallskip}\hline\noalign{\smallskip}
%Baseline (Culotta4) &  1,812,682 & 94.85 \\
%Signorini4 & 1,294,459 & 94.73 \\ 
%Chew3 & 307,884 & 94.48\\ 
%Google Flu Trends	& NA & \textit{99.12} \\ 
%\noalign{\smallskip}\hline\noalign{\smallskip}
Syndromes + Extra terms & 887,718 & 95.78 \\
%Negation & xxx & xxx \\
%Emoticon & xxx & xxx \\
%HashTags & xxx & xxx \\
%Humor & xxx & xxx \\
\textbf{Geo} & \textbf{213,989} & \textbf{96.03} \\
%Negation + Emoticon & xxx & xxx \\
%Negation + HashTags & xxx & xxx \\
%Negation + Humor & xxx & xxx \\
%Negation + Emoticon + HashTags + Humor & xxx & xxx \\
%\textbf{Negation + Emoticon + HashTags + Humor + Geo} & xxx & \textbf{xxx} \\
\noalign{\smallskip}\hline
\end{tabular}
\label{tab:sem2}
\end{table*}

\begin{table*}[tb]
\centering
\caption{\textbf{Types of tweets that are kept after semantic-based filtering of individual reports.}}
\begin{tabular}{ll}
\hline\noalign{\smallskip}
Types	& Tweet samples \\  
\noalign{\smallskip}\hline\noalign{\smallskip}
Influenza & I got flu n coughed a lot. Now my voice is like monster's voice. Rrr~ \\
confirmation  & \\
& @lisarob54 thanks lisa. He's got flu, bad head, aching all over, rotten  \\
& cough. Hasn't started making piggy noises yet tho!!! \\
& \\
& Barber just coughed on me in the chair. Pretty sure I now have swine flu \\ \noalign{\smallskip}\hline\noalign{\smallskip}
Influenza  & Are a sore throat and backache signs out the swine flu? Do I need to call \\
symptoms & the death squad? \\ 
& \\
& My day: flu-like symptoms (headache, body aches, cough, chills, 100.9 fever).\\ 
& Swine flu not ruled out. \#H1N1 \\ 
& \\
& Wish I could stay \& chat (or write a blog post) butI think I have the flu: 
\\ 
& aches, chills, fever, sore throat - it ain't purty. off to bed. \\ 
\noalign{\smallskip}\hline\noalign{\smallskip}
Flu shots & I'm still getting flu shots, nothing is worth flu turning into bronchitis \\
& into pneumonia \\ 
& \\
& I got a flu shot was the sickest I ever was . . .ended up with  bronchitis and \\
& worse \#ecowed \\ 
\\
& @pauloflaherty I'm home with bronchitis. Doc did strep test and flu test as  \\ 
& well. The flu test was torture! \\ 
\noalign{\smallskip}\hline\noalign{\smallskip}
Self protection & Cover your mouth if coughing, use a tissue, wash your hands often \& get a \\
&  flu shot - protect and defend your community from \#H1N1 \\
& \\
& 5 tips 2 keep you safe from the flu; When sick, stay home. Get flu shot. Hand\\ 
& wash/sanitize. Cover cough/sneeze. Healthy eating \& exercise. \\
&\\
& hands are dry so applied lotion, lotion is scented so nose runs, kleenex\\
& to * runny nose, wash hands bc of swine flu then have dry hands again \\ \noalign{\smallskip}\hline\noalign{\smallskip}
Medication & Wondering why I didn't take the flu shot, laying in bed with cough\\ 
& drops, medicine, and the remote  \\
&\\ 
& ... Cough medicine then sleep ... God protect me from the pig flu ... Plz and\\ 
& thnx ... K g'night twatters \\ 
\hline\noalign{\smallskip}
\end{tabular}
\label{tab:tweet-kept}
\end{table*}

Geographical features shown in Table~\ref{tab:sem} and Table~\ref{tab:sem2} helped increase correlation coefficients from 97.52\% to 98.39\% for \textit{Syndrome + ``flu''- URL} and 95.78\% to 96.03\% for \textit{Syndromes + Extra terms} but not statistically significant (p=0.18 and 0.45, respectively). We applied to filter by semantic features for \textit{Syndromes + Extra terms}, but we could not see any improvement, except for \textit{Geo} which helped increase the correlation coefficient from 95.78\% into 96.03\% but not significant  (p=0.45). This yields that geographical features might be more important than semantic features. 
Note that all p-values for differences take into account the number of tweets in each set.

When combining all semantic features, the best correlation coefficient was 98.46\% (\textit{Negation + Humor + Emoticon + HashTags + Geo}), an improvement of 3.98\% compared to \textit{Culotta4}. 

%\begin{figure}[t]
%\centering
%\includegraphics[scale=0.55]{./Fig5.eps}
%\caption{Percentage reduction in corpus size using semantic-based filtering methods on individual %reports (\textit{Syndrome + ``flu" - URL)}. The size of individual report corpus was 8,485 messages. %The original corpus size of Twitter data was 587,290,394 messages.}
%\label{fig:reduction}
%\end{figure}

% Reduction of corpus
%We note that whilst \textit{Cullota4} had over 1.8 million tweets, our \textit{Syndrome + ``flu" - URL)}  (individual reports) kept only %8,485 tweets (about 0.47\% and 0.0014\% of the total number of tweets in the \textit{Cullota4} data set and in the whole corpus, %respectively), but achieved a higher correlation coefficient (97.52\% vs. 94.85\%, p=0.06). Note that all p-values for differences take %into account the number of tweets in each set.
%The percentage reduction in corpus size incurred by semantic-based filtering methods on individual reports (\textit{Syndrome + ``flu" %- URL)} is shown in Figure~\ref{fig:reduction}.

\begin{figure}[t]
\centering
\includegraphics[scale=0.5]{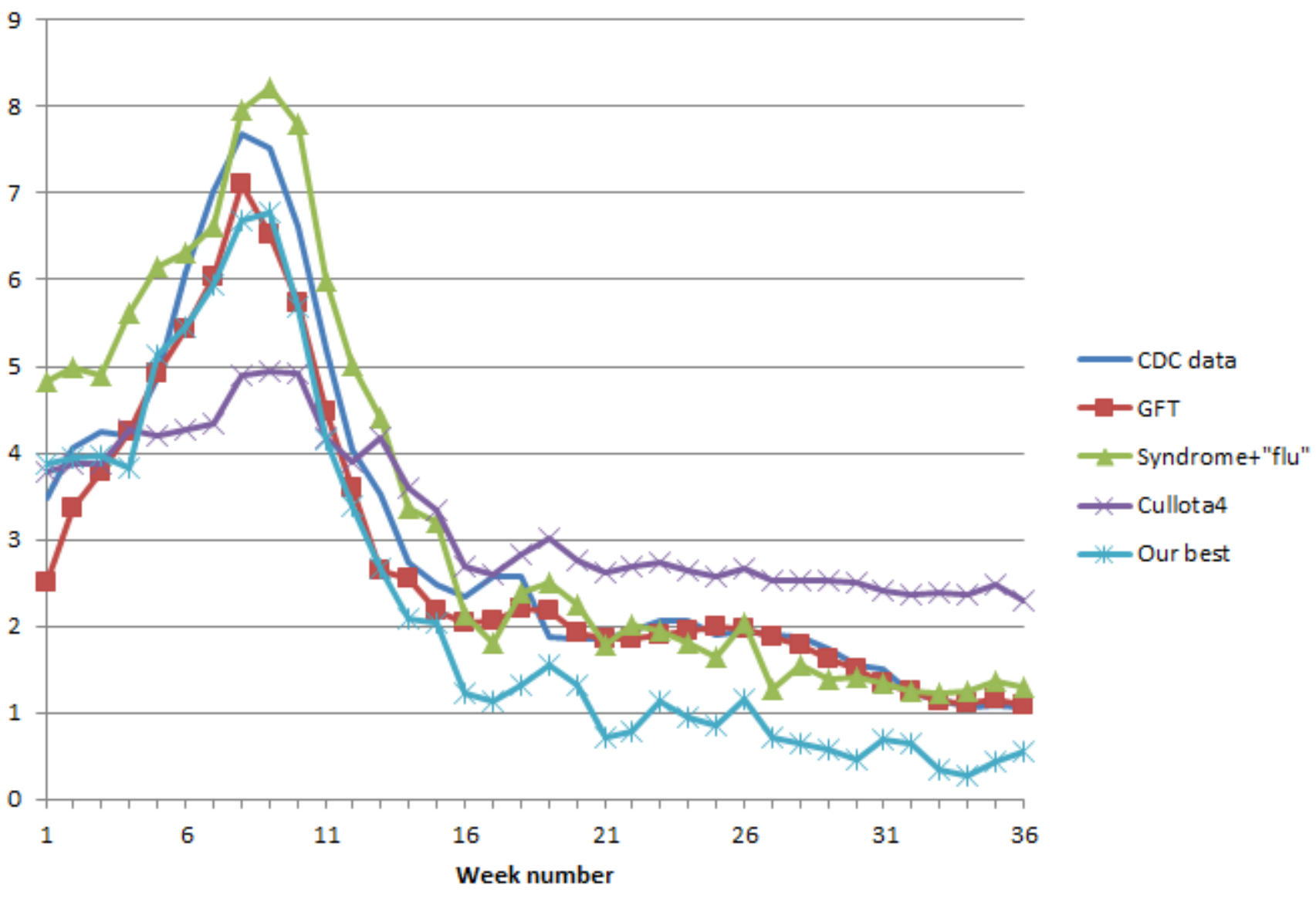}
\caption{Normalized Twitter data of \textit{Syndrome+``flu''}, the Best Combination (\textit{Negation + Humor + Emoticons + HashTags+ Geo}), Baseline (\textit{Culotta4}), Google Flu Trend (GFT) (for reference only, as it uses a different type of data), and CDC data (gold-standard). Y axis represents the ILI rate value (\%), values for GFT represent Google Flu Trends, and CDC data represents CDC's U.S. Outpatient Influenza-like Illness Surveillance Network. Other values are normalized for the number of tweets per week that correspond to each filtering method, i.e., $K\times$(number of tweets in the filtering method)/(total number of tweets that week), where $K=10^4,10^6,10^7$ for \textit{Cullota4}, \textit{Syndrome+``flu"}, and Best Combination, respectively.}
\label{fig:corre}
\end{figure}

For comparison, normalized Twitter data of \textit{Syndrome+``flu''}, Best Combination, i.e.\textit{ Negation + Humor + Emoticons + HashTags+ Geo}, Baseline (\textit{Culotta4}),  GFT and CDC data are shown in Figure~\ref{fig:corre}. 

\section{Discussion}
\label{s:discuss}

We observed from Table~\ref{tab:keyword} that the knowledge-based approach outperformed the empirical approach (97.52\% correlation coefficient for\textit{Syndrome + ``flu''- URL}, compared to 94.85\% for \textit{Culotta4}, p=0.06). As discussed in the Introduction, choosing keywords for filtering is very important and requires complex modeling to construct appropriate lists. The main difference between the knowledge-based approach and \textit{Culotta4} is that the knowledge-based approach first filtered tweets based on syndrome keywords and then filtered influenza from syndrome cases, while \textit{Culotta4} simply filtered using any of four keywords, i.e., ``flu'', ``headache'', ``sore throat'', and ``cough''. On the other hand, our method can obtain tweets that contain information on the combination of syndrome and influenza, thus yielding higher correlation with CDC rates. For example, \textit{Syndrome + ``flu''- URL} can contain tweets such as ``Down with a flu. Sore throat. Its burning like hale'' which requires a respiratory syndrome keyword in the BCO ontology (``sore throat'') first and then the keyword ``flu''.  \textit{Culotta4} had false positive in messages such as ``Guh, lack of sleep and a whole lot of algebra homework makes for one hell of a headache" or ``Has a wicked headache, as well as a new bill to pay'' which have no relation to influenza. 

Table~\ref{tab:keyword} also shows that individual reports had a positive effect on the ILI rate. Individual or personal reports, i.e., those where the subject of the message was the author, achieved a correlation coefficient of 97.52\% (\textit{Syndrome+``flu''}), compared to 97.12\% of comments on external events (\textit{Syndrome+``flu''-URL}) (p=0.38). We believe that this may have happened because the ILI rate is reflected in reports of individual and specific cases, but may not be reflected in general cases or undetermined groups. Therefore, individual reports seem more helpful to improve the ILI rate. For example, a tweet ``bad cough bad flu '' which implicitly indicates that the user has a ILI  would be more useful than an event-based tweet such as ``7-year-old boy dies of flu, pneumonia $<$URL$>$''.

Table~\ref{tab:sem} and Table~\ref{tab:sem2} show that the geographic features help to significantly improve performance in comparison to semantic features. In \textit{Culotta4}, it is implicitly assumed that the ILI rate in the US can be calculated based on Twitter messages from worldwide Twitter data. However, in reality, the ILI rate varies across administrative regions, including nation states. Moreover, the number of new Twitter users has increased dramatically daily with a number of 300,000 per day based on statistics for April 2010 \cite{chirp2010}. Geography was an important feature to track with the ILI rate in our data set. In retrospect, we could explain why features such as negation did not help, and could have even harmed the correlation: even if individuals did not have ILI, they were aware of it and could be commenting on close friends or family, who might not be tweeting about their own ailment. The same applies for Humor, Emoticons (which could represent sympathy messages), and HashTags.

Table~\ref{tab:tweet-kept} shows types of tweets after semantic-based filtering with individual reports. We classified the classified tweets into five types: influenza confirmation, influenza symptoms, flu shots, self protection, and medication. Among them, influenza confirmation and influenza symptoms were the most frequent.

There were several difficult cases in our method that need to be filtered out. 
For example, tweets about flu related opinions such as ``Just boarded my flight--hearing coughs, sneezes, sniffles. Gross. Maybe this us how I got a cold. Cold/flu season is here.''. Difficult cases also include tweets about flu vaccination and other prevention measures, e.g., ``is now properly vaccinated against the flu and pneumonia...oh and has a TB skin test..hmmm it is turning more red than usual'', or information about flu symptoms such as ``flu symptoms seem to have a cascade order - starts with 1.sore throat then 2.fever 3.aches\&pains 4.coughing''. Such messages are still not filtered out using our method. These cases may need to be handled by advanced natural language processing methods.

We compared our results to Google Flu Trends (GFT) over the same period. The correlation coefficient of GFT to CDC data was 99.12\%, higher than our best results, 98.46\% (\textit{Negation + Emoticon + HashTags + Humor + Geo}) but not statistically significant (p=0.13). We think one of the reasons for the difference between GFT and our methods relates to the quality of the data. GFT can access the whole query log of users from the Google search engine, while our data is a subset of the total Twitter data, with an unknown sampling rate. As we do not know the sampling rate precisely, we speculate that if the sampling rate were higher, the correlation coefficient could get closer to that of GFT. Furthermore, GFT can accurately identify geographical location by looking up the static IP address of machines, and Twitter's profile does not reveal IP address, but primarily free text submitted by the users. However, Twitter provides API functions at \url{https://dev.twitter.com} that can classify tweets by geolocations through latitude/longitude or city/province names. For example, Flu Detector retrieved tweets through this way by limiting to 49 urban centres in the UK only \cite{lampos2010}. Moreover, Twitter data are public and accessible freely through the Twitter API function, while users query data from Google users is closed and cannot typically be accessed by the research community. 

There are still several remaining avenues to improve the methods presented here. The first is establishing the scope of negation, which plays an important role in sentiment analysis \cite{Wiegand2010}, especially when there are more than two negations within a tweet. For example, ``yes i would really like to stop coughing. no i do not have swine flu''. The second is determining whether the tweet is factive or modal, reflecting user levels of belief in a particular condition, e.g., ``got cold but not sure I got flu''. The third avenue is to develop an explicit linkage with sentiment analysis and opinion mining \cite{pak10}. For example, determine the relation user responses to influenza news happening elsewhere and the ILI rate. Additionally, deeper analyses of demographics such as age, gender as well as user preferences, which are available in user profiles in Twitter could reveal important clusters. Such data are not available in Google Flu Trends, and could be an important enhancement for flu risk analysis.

\section{Conclusion}
\label{s:conclusion}

This study has shown that Twitter messages can be used to track the ILI rate with a high degree of correlation to official government data. By analyzing Twitter messages, we showed that the use of keywords based on a knowledge-based approach is beneficial. Furthermore, the use of semantic feature filtering was shown to be useful for selecting tweets based on geography. The approach is systematic and general, so it may be applicable to a wide range of diseases and syndromes. Further investigation in this direction is warranted. Other fruitful areas for future study include detection of predictive signals, and integration of data signals from social and news media. 
We implemented an experimental system called Dizie (Disease Information Extraction), available at \url{http://born.nii.ac.jp/dizieproj/_dev/}, to track ILI and five other syndromes using the methods and insights from this study.

\section*{Acknowledgements} 
We would like to thank Brendan O'Connor in Carnegie Mellon University for his supports in providing Twitter data, Aron Cullota in Southwestern Louisiana University for valuable comments and discussions. The first and third authors were partially supported by grant-in-aid funding from the National Institute of Informatics, Japan.

% BibTeX users please use one of
%\bibliographystyle{spbasic}      % basic style, author-year citations
%\bibliographystyle{spmpsci}      % mathematics and physical sciences
%\bibliographystyle{spphys}       % APS-like style for physics

% references section
% can use a bibliography generated by BibTeX as a .bbl file
% BibTeX documentation can be easily obtained at:
% http://www.ctan.org/tex-archive/biblio/bibtex/contrib/doc/
% The IEEEtran BibTeX style support page is at:
% http://www.michaelshell.org/tex/ieeetran/bibtex/
\bibliographystyle{IEEEtran}
% argument is your BibTeX string definitions and bibliography database(s)
%\bibliography{IEEEabrv,../bib/paper}
%
% <OR> manually copy in the resultant .bbl file
% set second argument of \begin to the number of references
% (used to reserve space for the reference number labels box)

%\bibliography{../../../../Reference/reference}
\bibliography{./reference}

% that's all folks
\end{document}